%% file: paper.tex
\pdfoutput=1
\documentclass[conference]{IEEEtran}
\IEEEoverridecommandlockouts

\usepackage{cite}
\usepackage[dvipsnames]{xcolor}
\usepackage{amsmath,amssymb,amsfonts}
\usepackage{textcomp}

\usepackage{enumitem}
\usepackage{verbatim}
\usepackage{beramono}
\usepackage{graphicx}
\usepackage{tikz}
\usetikzlibrary{arrows.meta,calc}
\usepackage{fix-cm}
\usepackage{tgheros}
\usepackage{sansmath}
\DeclareMathSizes{6.5}{6.5}{6}{6}
\usepackage{booktabs}
\usepackage{multirow}
\usepackage[export]{adjustbox}
\usepackage{tablefootnote}
\usepackage[]{hyperref}
\usepackage{microtype}
\usepackage{array}
\usepackage{footnote}
\usepackage{tabularx}
\usepackage{makecell}
\usepackage[utf8]{inputenc}
\usepackage{amsthm}

\makesavenoteenv{tabular}
\makesavenoteenv{table}

\usepackage[T1]{fontenc}

\hypersetup{hidelinks}
\AtBeginDocument{%
  \def\sectionautorefname{Section}%
  \def\subsectionautorefname{Section}%
  \def\subsubsectionautorefname{Section}%
  \def\appendixautorefname{Appendix}%
  \def\equationautorefname~#1\null{Eq.~(#1)\null}%
}
\newcommand{\appref}[1]{{\def\sectionautorefname{\appendixautorefname}%
  \let\subsectionautorefname\sectionautorefname
  \let\subsubsectionautorefname\sectionautorefname\autoref{#1}}}

\newcommand{\paratitle}[1]{\textbf{#1.}\enspace}

\def\BibTeX{{\rm B\kern-.05em{\sc i\kern-.025em b}\kern-.08em
    T\kern-.1667em\lower.7ex\hbox{E}\kern-.125emX}}
\begin{document}

\title{Bit-Flip Attacks on Quantized Neural Networks without Knowledge of Model Parameters}

\author{
\IEEEauthorblockN{Mahmoud Abumandour}
\IEEEauthorblockA{Simon Fraser University\\
Burnaby, BC, Canada\\
mahmoud\_abumandour@sfu.ca}
\and
\IEEEauthorblockN{Srinija Ramichetty}
\IEEEauthorblockA{George Washington University\\
Washington, DC, USA\\
sramichetty9@gwmail.gwu.edu}
\and
\IEEEauthorblockN{Guru Venkataramani}
\IEEEauthorblockA{George Washington University\\
Washington, DC, USA\\
guruv@gwu.edu}
\and
\IEEEauthorblockN{Alaa R. Alameldeen}
\IEEEauthorblockA{Simon Fraser University\\
Burnaby, BC, Canada\\
alaa@cs.sfu.ca}
}

\maketitle

\begin{abstract}
Model quantization improves the efficiency of deep neural networks (DNNs) and their resilience to random bit flips. However, quantized DNNs retain small high-precision tensors for numerical stability. We show that this quantization asymmetry concentrates bit-flip vulnerability in these tensors. We propose CoarseHammer, a data-independent attack that accepts any target bit without knowing parameter values, gradients, or data. Under ideal fault placement, CoarseHammer reduces six CNNs and four LLMs to random-guess accuracy in 81.5–100\% of runs, with median successful runs requiring at most 24 flips. On DRAM models with ultra-low Rowhammer vulnerability (0.01 flippable bits per million), CoarseHammer outperforms BFA, DeepHammer, and B3FA across six evaluated models, succeeding in up to 84\% of runs while none of the baselines succeed. On a real system, despite address space layout randomization and without the model file, CoarseHammer brings the median accuracy of all ten models within 2.2\% of random-guess accuracy.
\end{abstract}


\section{Introduction}

Model quantization improves the efficiency of deep neural networks (DNNs) and is widely regarded as improving their resilience to random bit flips, since a lower-precision parameter limits how far one flip can move it~\cite{hong2019terminal, yao_2020_deephammer, zhang2025verificationbitflipattacksquantized, 9156736}. However, quantized DNNs are rarely uniformly quantized. For numerical stability and accuracy preservation, inference frameworks retain a small fraction of a model's parameters at high precision, such as the biases of convolutional neural networks (CNNs) and the normalization weights of large language models (LLMs)~\cite{8578384, rotem2018glow, nvidia_tensorrt_llm_2026, tvm}. We show that this asymmetry creates an uneven distribution of bit-fault vulnerability.

A bit-flip attack (BFA) degrades a DNN by corrupting a few stored parameters the attacker chooses, rather than random ones~\cite{hong2019terminal, rakin2019bit, yao_2020_deephammer}. On shared inference platforms and edge devices, an unprivileged process can mount a BFA with Rowhammer, which repeatedly activates DRAM rows until bits in nearby rows flip~\cite{6853210, yao_2020_deephammer}. Most BFAs on model parameters search for bit-flip chains that, when flipped, achieve the attacker's objective, and these bit-search attacks rely on three assumptions~\cite{rakin2019bit, yao_2020_deephammer, coalson2024prisonbreak}. The attacker must know the exact parameter values of the victim model and have a batch of test data, because the search computes gradients from both. Every bit in the chain must also reliably flip, which requires a vulnerable DRAM cell under each chosen bit. However, page placement is nondeterministic~\cite{rambleed}, and vulnerable cells are sparse~\cite{8835222, frodo_2022, pinpoint_2019}. Finally, the parameter values must stay fixed after the search, since the chain is computed from them and becomes stale once the victim fine-tunes the model on the data it serves~\cite{odl_2017, usenix2020_updates_leak}.

To find targets that do not require any of these assumptions, we measure the damage from a random bit flip in high-precision tensors and in the quantized tensors of six CNNs and four LLMs. In ResNet18, a single random bit flip in a high-precision tensor causes a median accuracy drop of $75$ percentage points, compared with $0\%$ for a flip in a quantized tensor. Flips in the high-precision tensors of the four LLMs likewise degrade the LLMs substantially more than flips in their quantized weights. The \texttt{int8} tensors hold up to $99.8\%$ of the CNNs' parameter bits, so a uniformly random flip almost always lands where it does little damage. Quantization therefore protects a model against flips spread uniformly across its parameter bits, whereas flips aimed at its high-precision tensors bypass this protection.

Using this insight, we propose \emph{CoarseHammer}, a data-independent BFA that targets the high-precision tensors by their bit width and cardinality, which induces flips at any of the bits of these tensors. An attacker can identify the targets statically from the victim's model architecture, which can be recovered stealthily from a co-located victim~\cite{usenix2020_cache_telepathy, own_nas, usenix25_neuroscope, 10.1145/3373376.3378460,usenix23_btd,usenix22_dnd}, or from the official query interface~\cite{pasquini2025llmmapfingerprintinglargelanguage, iourovitski2024hideseekfingerprintinglarge}. Selecting tensors by these properties removes the need to search for a chain of individual bits and avoids the three assumptions above. CoarseHammer does not require knowing parameter values, gradients, or test data, and it accepts any flip that lands in a target tensor. Subsequently, CoarseHammer also does not impose precision requirements on the vulnerable DRAM cells that it uses, unlike prior attacks~\cite{yao_2020_deephammer, rakin2021t, coalson2024prisonbreak, usenix24_onebf_matters, comp_models_built_in_expoloits_2025, iccd24_semiblackbox}. Its targets stay the same when the victim fine-tunes the model, because fine-tuning changes neither the quantization scheme nor the tensors' cardinality. Under ideal fault placement, CoarseHammer brings all ten models we test to random-guess accuracy with an attack success rate of $81.5$-$100\%$ with a median of at most 24 bit-flips. \autoref{tab:attack_comparison} compares CoarseHammer with prior BFAs.

Accepting any target bit also allows CoarseHammer to succeed on DRAM with sparse flippable cells that are unlikely to match a chain of exact bits. We analyze this flip placement advantage and compare CoarseHammer to two white-box attacks, BFA~\cite{rakin2019bit} and DeepHammer~\cite{yao_2020_deephammer}, in addition to B3FA~\cite{iccd24_semiblackbox} which needs only part of the parameters. On simulated DRAM models with ultra-low Rowhammer vulnerability, CoarseHammer reaches random-guess accuracy more often than the three attacks on every evaluated model and vulnerability rate. At $0.01$ flippable bits per million, none of the three attacks brings any model to random-guess accuracy, while CoarseHammer does so in up to $84\%$ of the runs.

We demonstrate that with the weights quantized to different bit widths, CoarseHammer succeeds in up to $98.7\%$ of runs. When the victim fine-tunes the six CNNs for $50$ epochs while CoarseHammer flips up to four bits per epoch, every run on four of them reaches random-guess accuracy, and the mean accuracy of ResNet18 and ResNet50 drops to $16.4\%$ and $24.6\%$ from over $94\%$. This shows that CoarseHammer does not depend on the initial or any intermediate parameter values, and does not depend on the bit-widths used to quantize the model.

We then evaluate CoarseHammer in the presence of two BFA defenses. Range restriction~\cite{9505066} leaves CoarseHammer's success rate unchanged on all six CNNs, because \texttt{int8} inference already clips activations to nearly the same bounds. Aegis~\cite{wang2023aegis} attaches internal classifiers to residual blocks of the model and answers each input from a randomly chosen exit, so a flip in one layer reaches only the exits after that layer. However, the tensors that CoarseHammer targets feed nearly every exit, so the exits fail together. CoarseHammer still brings ResNet18 to random-guess accuracy after a median of $6$ bit flips in every run with Aegis in place, compared to a median of $4$ bit flips without Aegis.

Finally, we run CoarseHammer end-to-end on a real system with in-DRAM Rowhammer mitigation, against a victim program that loads its model with PyTorch~\cite{NEURIPS2019_9015} into heap memory. Offline, the attacker \emph{templates} its own memory (i.e., it hammers it to find physical frames with flippable cells). Address space layout randomization (ASLR) changes which heap page holds each target tensor in each victim load. However, the loader allocates tensors in an order fixed by their shapes and types, so profiling a dummy model of the recovered architecture reveals the page faults at which target pages receive their frames. At those faults, CoarseHammer hands the victim its templated frames. Since a flip anywhere in a target tensor is useful, it hammers all of these frames in every load, without learning which target page landed on which frame. The median accuracy over all runs is at most $0.3$ points above random-guess accuracy for every model except ResNet50, which falls from $96.1\%$ to $13.4\%$ (random guess: $11.2\%$).

In summary, we make the following contributions:

\begin{itemize}
    \item We show that bit-flip damage in quantized DNNs concentrates in a few high-precision tensors, which an attacker can locate from their bit width and size.
    \item We propose CoarseHammer, a BFA that accepts a flip at any bit of these tensors and needs no weights, gradients, or data. On DRAM modules with $0.01$ flippable bits per million, it reaches random-guess accuracy in up to $84\%$ of runs, while BFA, DeepHammer, and B3FA never do.
    \item We demonstrate CoarseHammer end-to-end on a real system, where it steers the victim's target pages onto templated frames. CoarseHammer substantially degrades all the models we test to within $2\%$ of random-guess accuracy despite ASLR, without the model file or the parameter values.
    \item CoarseHammer remains effective with weights quantized different bit widths, where it succeeds in at least $94.6\%$ of runs, and under continued fine-tuning. Range restriction leaves CoarseHammer's success rate unchanged. Against the multi-exit defense Aegis, CoarseHammer still reaches random-guess accuracy in every run after a median of two additional flips.
\end{itemize}

\newcommand*\emptycirc[1][1ex]{\tikz[baseline=-0.6ex]\draw[thick] (0,0) circle (#1);}

\newcommand*\halfcirc[1][1ex]{%
  \begin{tikzpicture}[baseline=-0.6ex]
  \draw[fill] (0,0)-- (90:#1) arc (90:270:#1) -- cycle ;
  \draw (0,0) circle (#1);
  \end{tikzpicture}}

\newcommand*\fullcirc[1][1ex]{\tikz[baseline=-0.6ex]\fill (0,0) circle (#1);}

\newcommand{\cmark}{\fullcirc}
\newcommand{\xmark}{\emptycirc}
\newcommand{\pmark}{\halfcirc}

\begin{table}[t]
\centering
\caption{CoarseHammer against prior BFAs. \cmark[0.65ex]: yes,
\pmark[0.65ex]: partially, \xmark[0.65ex]: no. $^\dagger$ marks
attacks on code, the rest attack parameters.}
\small
\resizebox{8.5cm}{!}{%
\begin{tabular}{l||ccccc}
\toprule
\textbf{Attack} & \makecell{Requires \\ Arch.} & \makecell{Requires \\ Weights} & \makecell{Requires \\ Test Set} & \makecell{Exact \\ Bit Flips} & \makecell{Targets Global \\ Libraries}
\\
\midrule
BFA~\cite{rakin2019bit}                                    &  \cmark & \cmark & \pmark & \cmark & \xmark \\
DeepHammer~\cite{yao_2020_deephammer}                      &  \cmark & \cmark & \pmark & \cmark & \xmark \\
DeepVenom~\cite{cai_2024_deepvenom}                        &  \cmark & \cmark & \pmark & \cmark & \xmark \\
PrisonBreak~\cite{coalson2024prisonbreak}                  &  \cmark & \cmark & \pmark & \cmark & \xmark \\
TA-LBF~\cite{bai2021targeted}                              &  \cmark & \cmark & \pmark & \cmark & \xmark \\
ZeBRA~\cite{park2021zebra}                                 &  \cmark & \cmark & \xmark & \cmark & \xmark \\
B3FA~\cite{iccd24_semiblackbox}                            &  \pmark & \pmark & \xmark & \cmark & \xmark \\
Invisible Hands~\cite{almalky2025invisible}                &  \cmark & \pmark & \pmark & \cmark & \xmark \\
Chen et al.$^\dagger$~\cite{comp_models_built_in_expoloits_2025} &  \cmark & \xmark & \xmark & \cmark & \xmark \\
FrameFlip$^\dagger$~\cite{usenix24_onebf_matters}          &  \xmark & \xmark & \xmark & \cmark & \cmark \\
\midrule
\textbf{CoarseHammer (ours)} & \cmark & \xmark & \xmark & \xmark &  \xmark \\
\bottomrule
\end{tabular}
}
\label{tab:attack_comparison}
\end{table}

\section{Background}

\subsection{DNN Quantization}
\label{sec:dnn_quant}

Quantization defines how the stored bits of a parameter map to the real value the model computes with~\cite{8578384}. An integer encoding $z$ represents the parameter value $s(z-z_0)$, where $s$ is the scale and $z_0$ is the zero point. Symmetric quantization fixes $z_0 = 0$~\cite{DBLP:journals/corr/abs-2106-08295}. A single scale covers an entire tensor, a channel, or, in LLM weight formats, a block of consecutive integers~\cite{DBLP:journals/corr/abs-2106-08295, llamacpp}.

Integer-only inference keeps biases and accumulators at a higher bit width than the weights, which make up the vast majority of parameters~\cite{8578384, DBLP:journals/corr/abs-2106-08295}. A linear layer sums products of \texttt{int8} weights and activations in a 32-bit integer (\texttt{int32}) accumulator to avoid overflow. Each bias is stored as an \texttt{int32} code at the accumulator's scale and added without rescaling~\cite{8578384}. This wide format adds little storage because biases are a tiny fraction of the parameters. Preserving bias precision also avoids a systematic rounding offset in every output to which the bias is added~\cite{8578384, 37631}.

Widespread inference frameworks keep biases at this higher precision~\cite{litert_quant_spec, onnx_qlinearconv, rotem2018glow,tvm}. For instance, TFLite's integer-only inference folds batch normalization into preceding convolution layers in CNNs and stores the biases as \texttt{int32} codes~\cite{8578384}, and PyTorch~\cite{NEURIPS2019_9015} also keeps the biases of its quantized operators in either \texttt{int32} or \texttt{float32}. Quantized LLMs make the same exception, where Softmax, GELU and layer normalization (LayerNorm) are approximated in \texttt{int32}~\cite{kim2021bert, li2023vit} or kept in floating point~\cite{ShenDYMYGMK20, Bhandare2019Efficient8Q, nvidia_tensorrt_llm_2026}. The Official GGUF~\cite{gerganov2023gguf} model files of Llama 3.2~\cite{dubey2024llama3herdmodels} and Gemma 2~\cite{gemmateam2024gemma2improvingopen} store their RMSNorm weights in floating point~\cite{llamacpp}.

\subsection{Bit-Flip Attacks}
\label{sec:bg_bfa}
A targeted BFA makes the DNN mispredict \emph{specific inputs} while holding accuracy on the rest~\cite{rakin2021t, tol_2021_dont_knock}. An untargeted BFA instead degrades overall accuracy as far as a few flips allow~\cite{yao_2020_deephammer}.

Most prior BFAs against quantized models are white-box and require knowledge of the model architecture, the parameter values, and a set of test data~\cite{yao_2020_deephammer, rakin2019bit, coalson2024prisonbreak, bai2021targeted}. Using this information, a bit-search algorithm constructs a bit-flip chain to maximize the model's loss on the test set. At each step, the algorithm selects the bit of the stored parameters $\Theta$ whose flip maximizes the increase in $\mathcal{L}(f(x;\Theta))$, the loss of model $f$ on test inputs $x$~\cite{rakin2019bit}. Selecting that bit requires evaluating gradients at the true parameter values.

\subsection{Flipping Bits using Rowhammer}

DRAM stores bits as charge in memory cells organized into rows. Activating an \emph{aggressor} row repeatedly accelerates charge leakage in nearby \emph{victim} rows. Cells can flip when the activation count between refreshes exceeds a device-dependent \emph{Rowhammer threshold}~\cite{6853210}. The locations of vulnerable cells and their flip directions depend on the physical DRAM device, which limits fault injection to a sparse set of cells. Rowhammer attacks have leaked information and escalated privileges on embedded systems, mobile devices and servers~\cite{embeddedrowhammer, 10.1145/2976749.2978406, kaur_2023_flipping_bits_like_a_pro, 9229701, projectzero2015, 9152631}. Rowhammer is the main method used to realize BFAs against DNNs~\cite{yao_2020_deephammer, rakin2021t, coalson2024prisonbreak, usenix24_onebf_matters, comp_models_built_in_expoloits_2025, iccd24_semiblackbox}.

Attacks on DNNs must first \emph{template} DRAM to identify physical frames that contain flippable bit cells and the aggressor rows that can be hammered to induce those flips~\cite{jattke_2022_blacksmith}. We call the resulting profile a template. The attacker then steers victim data onto these frames by tricking the OS into placing it there, and hammers the corresponding aggressor rows~\cite{rambleed, yao_2020_deephammer}.

\section{Threat Model}
\label{sec:threat-model}

We define the attacker's capabilities, the victim's deployment assumptions, and the defenses present on the victim system.

\subsection{The Victim}

The target model is securely trained or obtained from a trusted source, without interference from an adversary through model poisoning~\cite{sp2018_poisining} or backdoor insertion~\cite{Liu2018TrojaningAO}. It is deployed on a resource-sharing platform such as a machine-learning-as-a-service cloud or an edge device, where multiple users share physical resources~\cite{yao_2020_deephammer, 197231, coalson2024prisonbreak, usenix2020_cache_telepathy, iccd24_semiblackbox, sagemaker, vertex, nvda_edgeai}. The inference stack, including the hardware, the kernel, and the inference framework, is uncompromised and up to date. We also assume the DRAM chips have the standard in-DRAM Rowhammer mitigation Targeted Row Refresh (TRR), which most DDR4 modules include~\cite{9152631}.

Deployed models experience distribution shift between their training data and real inputs~\cite{shift}. Some deployments fine-tune continually on the data they serve~\cite{odl_2017, usenix2020_updates_leak}, so the weights change in ways that the attacker does not know. We permit model fine-tuning and parameter changes during deployment.

\subsection{The Attacker}

An unprivileged attacker process is co-located with the victim process, which is feasible via standard co-location techniques from prior work~\cite{Varadarajan15, Atya19, Atya17, Ristenpart09, Xu15, Zhang14}. It has \emph{no knowledge} of the model’s parameter values, training data, hyperparameters, or gradients, and reaches the model only through its query interface.

We assume the attacker recovers the victim's operator graph and tensor shapes. Prior work recovers it through side channels from a co-located process~\cite{usenix2020_cache_telepathy, own_nas} or with physical access~\cite{10.1145/3373376.3378460}, from compiled model executables~\cite{usenix23_btd, usenix22_dnd, usenix25_neuroscope}, and for LLMs through the query interface~\cite{pasquini2025llmmapfingerprintinglargelanguage, iourovitski2024hideseekfingerprintinglarge}.

The attacker injects faults with Rowhammer~\cite{6853210}. On modules with in-DRAM mitigations, this requires circumventing them and recovering the DRAM undocumented mapping from physical addresses to DRAM channel, rank, bank, and row, which the attacker needs to choose adjacent aggressor addresses next to the victim in the same bank. Prior work demonstrates both capabilities from unprivileged processes~\cite{9152631, jattke_2022_blacksmith, 197211}.

\section{Finding the Most Vulnerable Tensors}
\label{sec:measurement}

Bit-flip damage concentrates in the small high-precision tensors of the six CNNs in \autoref{tab:models}. For each parameter tensor, we flip one uniformly random bit and evaluate the model and average the squared change in pre-softmax logits, $E[\|\Delta f(x)\|_2^2]$, over $200$ fault trials. Across the six CNNs, a flip in the median bias tensor causes $65$-$12000\times$ the expected squared logit change of a flip in the median quantized weight tensor. Next, we measure how far corruption from these flips spreads through the model.

\begin{table}[t]
\centering
\caption{The ten models we evaluate CoarseHammer against, their datasets, parameter counts, pristine accuracy, and random-guess accuracy on the evaluation examples. CNNs are quantized to \texttt{int8}, while LLMs store their weights in the $8$-bit \texttt{Q8\_0} format~\cite{llamacpp}.}
\label{tab:models}
\resizebox{\columnwidth}{!}{%
\begin{tabular}{l||lrrr}
\toprule
\textbf{Model} & \textbf{Dataset} & \textbf{Params} & \multicolumn{2}{c}{\textbf{Accuracy (\%)}} \\
\cmidrule(lr){4-5}
 & & & Pristine & Random guess \\
\midrule
ResNet8~\cite{he2016deep}      & CIFAR-10~\cite{alexcifar10}            & $77.7$k & $78.0$ & $11.2$ \\
ResNet18     & CIFAR-10                               & $11.2$M & $95.9$ & $11.2$ \\
ResNet50     & CIFAR-10                               & $23.5$M & $96.1$ & $11.2$ \\
DenseNet121~\cite{huang2017densely}  & CIFAR-10                               & $7.2$M  & $82.2$ & $11.2$ \\
MobileNetV1~\cite{howard2017mobilenets}  & Visual Wake Words~\cite{chowdhery2019vww} & $211$k  & $85.4$ & $50.0$ \\
DS-CNN~\cite{zhang2018helloedgekeywordspotting}       & Speech Commands~\cite{warden2018speech} & $22.6$k & $92.0$ & $9.6$ \\
\midrule
Llama-3.2-3B~\cite{dubey2024llama3herdmodels}                 & MMLU~\cite{hendrycks2021measuringmassivemultitasklanguage} & $3.2$B & $62.2$ & $25.0$ \\
Gemma-2-2B~\cite{gemmateam2024gemma2improvingopen}            & MMLU & $2.6$B & $56.5$ & $25.0$ \\
Qwen2.5-1.5B~\cite{qwen25technicalreport}                     & MMLU & $1.5$B & $60.4$ & $25.0$ \\
Phi-4-Mini~\cite{microsoft2025phi4minitechnicalreportcompact} & MMLU & $3.8$B & $63.3$ & $25.0$ \\
\bottomrule
\end{tabular}}
\end{table}

\begin{figure*}[t]
\centering
    \includegraphics[width=\linewidth]{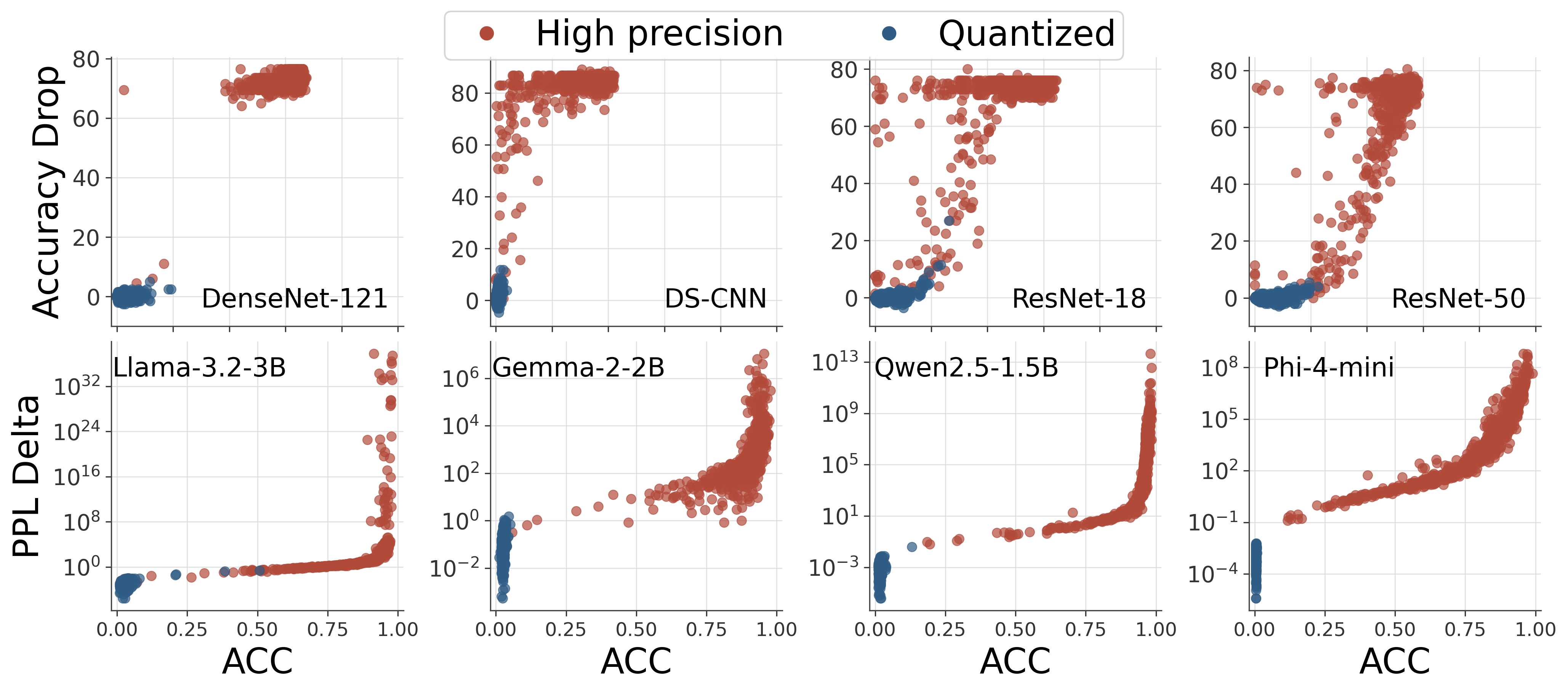}
\caption{Activation Change Coverage (ACC) at $\tau=10\%$ against degradation under random bit flips, showing the accuracy drop in percentage points for four CNNs (top), and the perplexity increase on WikiText for four LLMs (bottom). Each point is one fault trial with a single bit flip, injected into the high-precision tensors (red) or into the quantized tensors (blue).}
\label{fig:acc_8models}
\end{figure*}

\subsection{Activation Change Coverage}
\label{sec:acc}

Flips in high-precision tensors also spread further through the model. We measure whether such flips perturb a larger fraction of intermediate activations than a flip in the quantized weights, and whether this wider spread corresponds to greater task degradation.

To measure how far a corruption propagates through the model computation, we define \textit{Activation Change Coverage (ACC)} as the fraction of intermediate activations whose relative change from the original unmodified model exceeds a threshold $\tau$. For a model input $x$, let $a_i(x)$ and $a_i'(x)$ be the values of the $i$-th intermediate activation in the original and corrupted model, respectively; $N$ be the total number of measured activations, and $\epsilon$ be a small constant that keeps the ratio finite when $a_i(x)=0$. ACC is then given by \autoref{eqn:2}.

\begin{equation}
\mathrm{ACC}(x)=\frac{1}{N}\sum_{i=1}^{N}\mathbf{1}\left(\frac{|a_i'(x)-a_i(x)|}{|a_i(x)|+\epsilon}>\tau\right)
\label{eqn:2}
\end{equation}

\autoref{fig:acc_8models} plots ACC at $\tau = 10\%$ against model degradation for single random bit flips in four CNNs and four LLMs. For the CNNs, degradation is the drop in task accuracy. For the LLMs, it is the increase in perplexity (PPL) on a subset of the WikiText dataset, which measures how well the model predicts a sample text. In almost all cases, flips in high-precision tensors produce a higher ACC than flips in the quantized weights, and the higher ACC corresponds to larger degradation. In ResNet18, for example, the median ACC is $0.54$ for flips in high-precision tensors and $0.01$ for flips in quantized tensors, and the median accuracy drops are $75$ and $0$ percentage points, respectively.

ACC practically predicts the damage of a flip in a high-precision tensor more reliably than the damage of a flip in a quantized weight. In every CNN and LLM except DS-CNN, a flip in a high-precision tensor causes greater degradation when it affects a larger fraction of activations. In contrast, the rare damaging flips in quantized weights are not reliably the ones that spread furthest. In DS-CNN, high-precision flips are so often catastrophic that there is little variation in damage for ACC to predict. For an attacker, finding the most effective flips in the quantized weights therefore requires a search, as in prior attacks~\cite{yao_2020_deephammer,rakin2019bit,rakin2021t,iccd24_semiblackbox,coalson2024prisonbreak}. The greater damage from high-precision flips instead lets an attacker target a high-precision tensor without needing to identify its most damaging bits.

\section{The CoarseHammer Attack}
\label{sec:CoarseHammer_highlevel}
\label{sec:CoarseHammer_eval}

\begin{figure*}[t]
    \centering
    \input{figures/overview}
    \caption{Overview of CoarseHammer on ResNet8. Offline, architecture
    recovery yields the operator graph with bit widths, shapes and allocation
    order \textcircled{1}. A flip in a bias $T$ spreads through every
    later operator to the logits \textcircled{1=2}. Templating DRAM finds frames with
    flippable cells \textcircled{3}. Online, a planter thread pushes a
    templated frame $F_i$ and then $n{-}1$ filler frames onto the page frame
    cache (PFC) list, which is last-in, first-out; so the fillers occupy the
    victim's next $n{-}1$ faulted pages and $F_i$ occupies the $n$-th, which is the target page \textcircled{4}. The attacker then hammers the aggressor rows of $F_i$ \textcircled{5}, and
    accuracy falls to the random-guess level \textcircled{6}, shown as the mean of $1000$ ideal-placement runs.}
    \label{fig:overview}
\end{figure*}
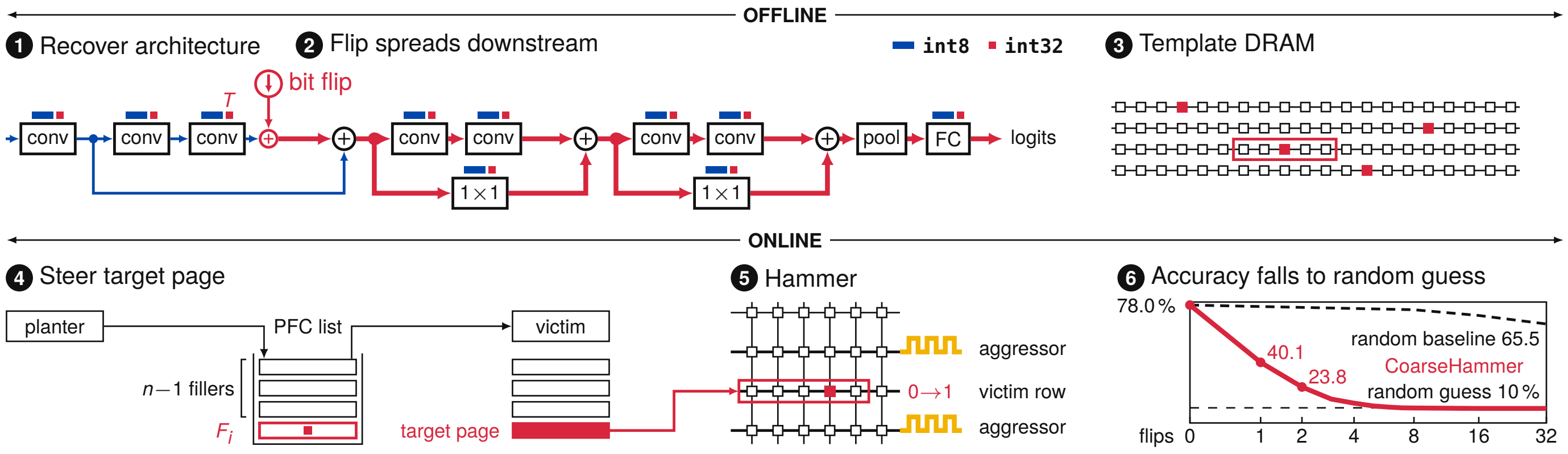

CoarseHammer attacks small high-precision tensors and accepts a flip at any of their bits, without reading parameter values or ranking bits by gradient. For CNNs, it ranks the \texttt{int32} tensors by increasing $n_T$, the number of elements in tensor $T$ divided by the number of operators that read it. This ranking favors small tensors and shared buffers, whose corruption reaches every operator that reads them. For LLMs, it targets normalization tensors and attention biases. \autoref{fig:overview} shows the end-to-end attack on ResNet8, and \autoref{sec:CoarseHammer_impl} describes how CoarseHammer finds, steers, and hammers the target pages on a real system.

\subsection{Realization}
\label{sec:CoarseHammer_impl}

On a real system, CoarseHammer runs in two phases. Offline, it locates the target pages and templates DRAM for frames with flippable cells. Online, it steers the victim's target pages onto templated frames and hammers the aggressor rows of those frames. Templating, address-mapping inference and hammering reuse published techniques and do not need privileges. DeepHammer placed weight pages through the last-in, first-out reuse of freed frames~\cite{yao_2020_deephammer}, and Rubicon controls this reuse at page granularity from an unprivileged process~\cite{bolcskei2025rubicon}. CoarseHammer's steering extends Frame Feng Shui~\cite{rambleed} to target pages deep in a model load, at positions that ASLR moves. We implement CoarseHammer against PyTorch~\cite{NEURIPS2019_9015} on Linux 7.0. In \autoref{sec:end_to_end}, we run it end-to-end on a real server. The target tensors are the biases, normalization and attention bias tensors of the models in \autoref{tab:models}.

\paratitle{Locating target pages}
The attacker recovers the victim architecture by matching a trace of operator calls and their latencies against offline profiles~\cite{own_nas}. It then loads a dummy model with the recovered shapes and random weights in its own copy of the victim's program. Since only the shapes and types of the tensors determine the allocation order, profiling runs of the dummy model show at which victim page fault each target page receives its frame.

\paratitle{Templating DRAM}
The attacker recovers the physical-to-DRAM address mapping from row-buffer conflicts~\cite{197211} and templates its own memory with the black-box fuzzer Blacksmith~\cite{jattke_2022_blacksmith}, whose hammering patterns circumvent TRR. For each frame $F_i$ with a flippable cell, it records the aggressor rows $A_i$ and the pattern that flipped the cell.

\paratitle{Steering target pages}
Online, CoarseHammer influences the kernel to back the victim's target pages with templated frames through the page frame cache (PFC), as in Frame Feng Shui~\cite{rambleed}. Linux serves single-page allocations from the PFC, a per-CPU list of free frames that it reuses in last-in, first-out order. An attacker and a victim on the same CPU therefore receive frames for their anonymous pages from the same list. To back the victim's $n$-th faulted page with a templated frame $F_i$, the attacker frees $F_i$ and then $n-1$ filler frames onto that list. The next $n-1$ page faults consume the filler frames, leaving $F_i$ for the target page. We call this sequence of freed frames a \emph{stack}. A frame's \emph{position} counts from the last frame freed, so $F_i$ sits at position $n$. The CNNs' bias pages come $14{,}700$ or more faults after \texttt{execve}, and the last normalization page of Gemma-2-2B comes about $694{,}000$ faults after \texttt{execve}. To avoid building stacks that span all these faults, CoarseHammer's planter thread pushes each stack shortly before the victim faults the target pages as described next.

\begin{figure*}[t]
\centering
\includegraphics[width=\textwidth]{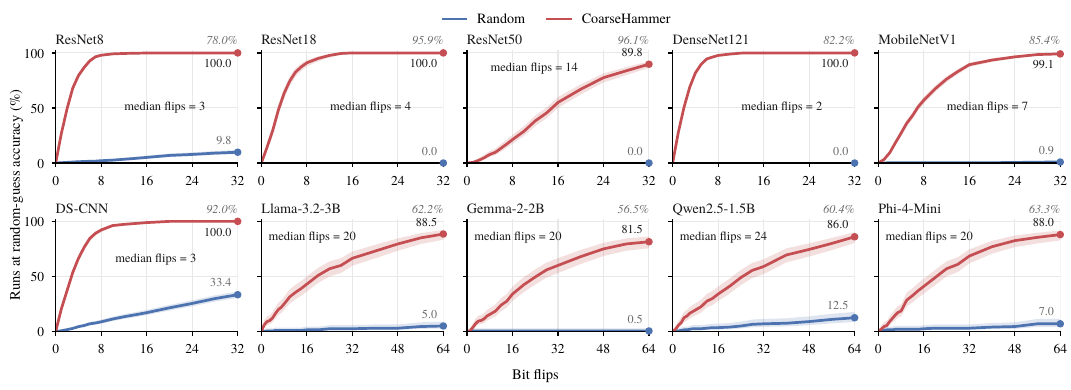}
\caption{Fraction of runs that have reached random-guess accuracy by each number of
flips, over $1000$ runs per model. The value at the flip
limit, $32$ for a CNN and $64$ for an LLM, is the attack success rate. Each
panel gives CoarseHammer's median flips to random-guess accuracy among its successful runs,
and the gray percentage beside each model name is its pristine accuracy from
\autoref{tab:models}. Bands are $95\%$ confidence intervals.}
\label{fig:attack_placement}
\end{figure*}

\paratitle{Synchronizing with the victim}
Since the victim's $n$-th fault after a push takes the frame at position $n$, each push must occur at the same point in every load. From another CPU, the attacker polls the victim's virtual size and minor-fault count in the world-readable \texttt{/proc/<pid>/stat}, and wakes the planter when the victim reaches this point. The planter sleeps on the victim's CPU and, once woken, must preempt the victim promptly, because the victim keeps faulting pages until the push. Under Linux's EEVDF scheduler~\cite{linux_eevdf}, a thread with the shorter time slice preempts the running thread, so the planter shortens its slice below the victim's, which requires no privileges. The Linux 7.0 kernel preempts lazily, so the planter runs at the victim's next return to user mode or the next timer tick.

For the CNNs, the first push must come early, while PyTorch's loader compiles the model code stored in the model file. Compilation grows the heap by hundreds of pages for temporary objects and then frees most of these objects, but the heap pages stay mapped and keep their frames. Bias tensors that fit into this freed memory are later allocated there, so their pages already hold frames from the compilation. The attacker therefore wakes the planter once the victim's virtual size reaches a value observed early in the compilation during the dummy model's profiling runs. The virtual size changes only when the victim grows its heap or maps memory, which the victim does in the same order in every load, so this value marks the same point of each load. In our experiments, one stack pushed at this point puts every bias page in the compiler's freed memory within its top $1{,}720$ positions, whereas these pages come $14{,}700$ or more faults after \texttt{execve}.

\paratitle{Targets that extend the heap}
Target tensors that do not fit into the heap memory that the compiler freed extend the heap top while the loader reads the weights. These are the bias tensors of the later layers of ResNet50 and DenseNet121 and most normalization tensors of the four LLMs. The loader reads each tensor with one \texttt{read()} call, so such a target faults its pages right after the \texttt{read()} of the preceding tensor returns. CoarseHammer therefore divides this part of the load into stages. For each stage, the attacker follows the victim's minor-fault count. It wakes the planter thread shortly before the count at which the profiling runs saw that \texttt{read()} end, so the planter pushes the stage's stack as the call returns.

\paratitle{Handling ASLR}
ASLR changes which tensors fit into the heap memory freed by the compiler and which heap pages hold the target tensors. The compiler faults the same heap pages at the same stack positions in every load, but ASLR changes which of those pages the target tensors later reuse. Target pages can therefore occupy different positions from load to load within a fixed set of candidate positions. Over $40$ profiling runs, ResNet18's bias pages occupied $79$ distinct stack positions. The interval from the earliest to the latest of those positions contained $262$ positions in total. The attacker places templated frames at positions that held target pages during profiling and filler frames at the other positions.

\paratitle{Hammering}
Once the victim loads its model, the attacker hammers the aggressor rows $A_i$ of every frame $F_i$ it placed, using the pattern that flipped the cell during memory templating. It hammers every placed frame because it does not learn which victim page received which frame.

\subsection{Experimental Setup}
\label{sec:exp_setup}

\paratitle{Models and datasets}
We evaluate CoarseHammer using the ten models in \autoref{tab:models}. We score each CNN on the same $1000$ examples of its dataset, and each LLM on
a fixed subset of $1000$ questions drawn from the MMLU test split. We count a run as successful once
the model's accuracy is at or below \textit{random-guess accuracy}.\footnote{We take random-guess accuracy to be the accuracy of always predicting the majority class, the class most common among the evaluation examples.} We draw the
CIFAR-10 and Speech Commands examples at random, so their classes are not
balanced, and random-guess accuracy on the testing subset is $11.2\%$ and
$9.6\%$. Random-guess accuracy on Visual Wake Words is $50\%$, and on MMLU it is $25\%$.

\paratitle{Flip sampling}
Each run samples distinct bits uniformly, flips them cumulatively, and measures accuracy. The random baseline samples from every bit of every parameter tensor. CoarseHammer samples from every bit of the high-precision tensors. We sample uniformly because a Rowhammer flip lands wherever the victim row has a weak cell~\cite{8835222, pinpoint_2019, frodo_2022}. To separate the effect of the flips from their placement, we first assume that every sampled bit flips, which we call \emph{ideal placement}, and add DRAM constraints in \autoref{sec:CoarseHammer_vs_attacks}.

\subsection{Attack Success}
\label{sec:CoarseHammer_conventional}

\paratitle{Random baseline}
After up to $64$ flips spread uniformly over all parameter bits, the median accuracy remains virtually unchanged on most models. After these flips, the random baseline brings the four LLMs to random-guess accuracy in at most $12.5\%$ of runs. The \texttt{int8} tensors hold $90.3\%$-$99.8\%$ of the CNNs' parameter bits, and the \texttt{Q8\_0} weight tensors hold over $99.96\%$ of the LLMs' parameter bits. A flip in these quantized tensors does less damage than a flip in the high-precision tensors, by orders of magnitude in the CNNs' logits, as shown in \autoref{sec:measurement}. Almost every random flip therefore lands where it does insignificant damage. DS-CNN keeps $9.7\%$ of its parameter bits in \texttt{int32} tensors, a larger share than the other CNNs. Random flips therefore hit DS-CNN's high-precision tensors more often, and the random baseline brings DS-CNN to random-guess accuracy in $33.4\%$ of runs.

\paratitle{CoarseHammer}
Sampling the flips uniformly from the high-precision tensors makes the attack destructive on all ten models using a small number of bits. CoarseHammer brings the six CNNs to random-guess accuracy in at least $89.8\%$ of runs within $32$ flips, and the four LLMs in at least $81.5\%$ of runs within $64$ flips as shown in \autoref{fig:attack_placement}. LLMs require more flips as they store high-precision parameters in floating point format, which does not incur substantial damage when a mantissa bit is flipped. In the hardest CNN, ResNet50, the median successful run needed $14$ flips, and in the hardest LLM, Qwen2.5-1.5B, it needed $24$ flips.

\subsection{DRAM Vulnerability Requirements}
\label{sec:CoarseHammer_vs_attacks}

Ideal placement assumes that every sampled bit flips, whereas in real DRAM a bit is flipped only if the physical frame that holds the bit's page has a flippable cell at the bit's offset. We now model the probability that at least one flip lands on a target bit. A bit-search attack needs the exact bits of its chain to flip, whereas CoarseHammer accepts a flip at any target bit, and we show that this difference lets CoarseHammer land flips on far less vulnerable DRAM.

\begin{figure}[t]
\centering
\includegraphics[width=\columnwidth]{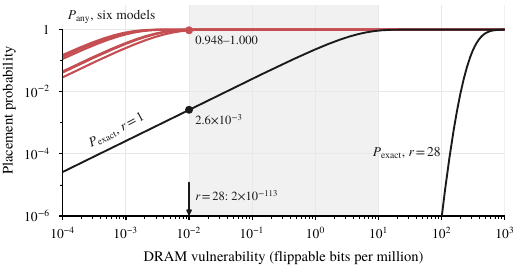}
\caption{Placement probability with $1$\,GiB templated. Red curves give
$P_{\mathrm{any}}$ for each of the six models, and black curves give
$P_{\mathrm{exact}}$ for $r$ exact bits. The gray band marks the
DRAM vulnerabilities of \autoref{fig:attack_comparison}.
}
\label{fig:placement_requirements}
\end{figure}

\begin{figure*}[t]
    \centering
    \includegraphics[width=\textwidth]{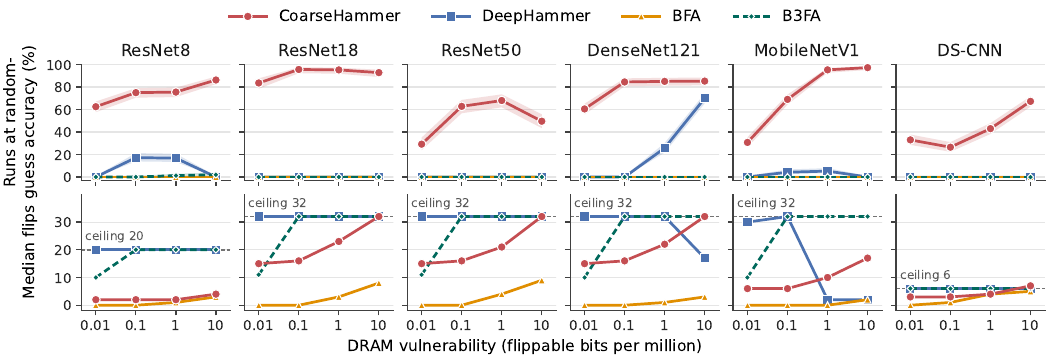}
    \caption{The four attacks on the same simulated templates of $1$\,GiB, one column per model.
    \emph{Top:} percentage of runs at random-guess accuracy after their last flip, over $400$ templates per point.
    \emph{Bottom:} median number of flips that land, at most $32$.}
    \label{fig:attack_comparison}
\end{figure*}

\paratitle{DRAM model}
The attacker templates $m$ physical frames of $4$\,KiB. We assume that each cell of the $m$ frames is flippable independently with the same probability $p$. The \emph{DRAM vulnerability} is the number of flippable bits per million. Hence, $p$ is the DRAM vulnerability divided by $10^6$. The victim stores its target bits in $C$ pages, which we call target pages. Since a frame can hold only one page, we split the $m$ frames evenly among the $C$ target pages, so each target page has its own $m/C$ candidate frames. After templating, the attacker places each target page on a candidate frame with a flippable cell under one of the page's target bits, if such a frame exists.

\paratitle{Any-bit placement}
CoarseHammer accepts a flip at any of the $L_c$ target bits in target page $c$. A single candidate frame has no flippable cell at any of these $L_c$ target-bit offsets with probability $(1-p)^{L_c}$. Across all $m/C$ candidate frames for page $c$, the probability that no frame can flip a target bit is therefore $(1-p)^{L_c m/C}$. Any-bit placement fails only if no target page has a candidate frame that can flip one of its target bits. Taking the complement of this failure probability gives~\autoref{eq:p-any}.

\begin{equation}
P_{\mathrm{any}}
  = 1 - \prod_{c}(1-p)^{L_c m/C}
  = 1 - (1-p)^{mL/C},
\label{eq:p-any}
\end{equation}

where $L=\sum_c L_c$ is the total number of target bits. For CoarseHammer, $L_c$ counts every bit of its target tensors in page $c$. Flipping bit $k$ of the \texttt{int32} code of bias $j$ with scale $s_j$ changes the parameter's real value by $\pm s_j * 2^{k}$, regardless of what value it stores, so CoarseHammer can accept a flip at any of these bits without knowing the weights.

\paratitle{Exact-bit placement}
A bit-search attack~\cite{yao_2020_deephammer,park2021zebra,rakin2021deepdup,comp_models_built_in_expoloits_2025,usenix24_onebf_matters,coalson2024prisonbreak,cai_2024_deepvenom,iccd24_semiblackbox,almalky2025invisible,bai2021targeted} instead needs $r$ exact bits to flip, and several of these bits can lie in one page. Let the $r$ bits occupy $g$ pages, with $r_c$ of them in page $c$, and split the $m$ frames evenly among the $g$ pages. Since a frame holds a single whole page, the bits of page $c$ land only on a frame with a flippable cell at each of their $r_c$ offsets, so at least one of the page's $m/g$ candidate frames has such cells with probability $1-(1-p^{r_c})^{m/g}$. All $g$ pages must land their bits, so exact-bit placement succeeds with the probability in~\autoref{eq:p-exact}, where the approximation holds for $mp^{r_c}/g\ll1$.

\begin{equation}
P_{\mathrm{exact}} = \prod_{c}\bigl(1-(1-p^{r_c})^{m/g}\bigr)\approx(m/g)^{g}\,p^{r}
\label{eq:p-exact}
\end{equation}

Since $(m/g)^{g}$ grows with $g$, the $r$ bits are most likely to land when they lie one per page, as in DeepHammer~\cite{yao_2020_deephammer}. We therefore evaluate this layout (which is the most favorable one for a bit-search attack), where $P_{\mathrm{exact}}\approx(mp/r)^{r}$. For one exact bit ($r=1$), each of the $m$ frames offers one chance to land the flip, at the bit's offset, so $P_{\mathrm{exact}}\approx mp$. Under any-bit placement, each candidate frame instead offers one chance per target bit of its page, so $P_{\mathrm{any}}\approx mpL/C$ when $mpL/C\ll1$. Any-bit placement is therefore about $L/C$ times as likely as placing one exact bit, where $L/C$ is the mean number of target bits per target page. With more exact bits, $P_{\mathrm{exact}}$ falls further below $P_{\mathrm{any}}$, because all $r$ bits must land and each gets only $m/r$ frames, as the black curves in \autoref{fig:placement_requirements} show.

As shown in~\autoref{sec:CoarseHammer_vs_attacks}, accepting any target bit makes at least one landed flip very likely even when the templated memory rarely provides a single exact bit. At $0.01$ flippable bits per million, any-bit placement succeeds with probability $0.948$-$1.000$ across the six models, whereas exact-bit placement succeeds with probability $2.6\times10^{-3}$ for one exact bit and about $2\times10^{-113}$ for $28$ exact bits (which corresponds to the mean number of flips that a white-box attack needs on a ResNet-20~\cite{9156736}). With these $28$ bits two per page, $P_{\mathrm{exact}}$ falls to about $7\times10^{-165}$. Since $P_{\mathrm{any}}$ grows with $p$, any-bit placement in $1$\,GiB succeeds with probability at least $0.948$ on each of the modules templated by Blacksmith~\cite{jattke_2022_blacksmith} under our placement model.

\subsection{Comparing CoarseHammer vs. Bit-Search Attacks}
\label{sec:CoarseHammer_vs_bitsearch}

We compare against BFA~\cite{rakin2019bit} and DeepHammer~\cite{yao_2020_deephammer}. Both pursue the same goal as CoarseHammer, but they read the weights and a set of test data. BFA introduced the progressive bit search, and DeepHammer is its Rowhammer realization. We also compare against B3FA~\cite{iccd24_semiblackbox}, the only BFA on model parameters in \autoref{tab:attack_comparison} that needs neither the full weights nor evaluation data. B3FA recovers part of the weight bits through a Rowhammer side channel, fills in the rest so that each weight is as close to zero as possible, and selects the sign bits of the largest weights in the convolution filters with the largest weight norm per element. We provide B3FA with $90\%$ of the weight bits, matching the highest recovery rate evaluated in~\cite{iccd24_semiblackbox}. BFA fixes its chain first and performs a flip only where the template has a flippable cell at that offset, so it keeps at most the first flip in its chain on each page. DeepHammer instead searches over the bits that the template can flip, and B3FA likewise ranks only the sign bits that the template can flip. BFA and DeepHammer stop once random-guess accuracy is reached; and B3FA, which reads no data, selects up to $32$ bits. With these flips, the BFA bit-chains bring ResNet8, ResNet18, DenseNet121, and DS-CNN to random-guess accuracy within $6-19$ flips. Under the same ideal-placement assumption, CoarseHammer's median successful run needs $2-4$ flips.

In this comparison, CoarseHammer targets all \texttt{int32} tensors, and any frame in the simulated $1$\,GiB template can hold any target page. For each target page, CoarseHammer scores every unused frame by summing the weights assigned to its flippable cells at offsets that hold target tensors. A flippable cell at an offset holding tensor $T$ contributes $n_{\min}/n_T$, where $n_T$ is the tensor's element count divided by the number of operators that read it. The normalization factor $n_{\min}$ is the smallest $n_T$ among the targets. CoarseHammer selects the frame with the highest score. This weighting favors flips in tensors with smaller $n_T$, so larger target tensors add candidate flips without taking priority over the smallest ones.

\paratitle{Baseline comparison}
CoarseHammer reaches random-guess accuracy more often than the three attacks on every model at every DRAM vulnerability we evaluate. At the lowest vulnerability, none of the other attacks reach random-guess accuracy on any model, while CoarseHammer does so in up to $84\%$ of runs. The bit searches fall short for different reasons. BFA fixes its chain before it sees the template, so few of its bits land, and it brings no model to random-guess accuracy at any vulnerability. DeepHammer and B3FA search over the bits that the template can flip, so their flips do land. At the two lowest vulnerabilities, DeepHammer lands more flips than CoarseHammer on every model, as shown in the bottom row of \autoref{fig:attack_comparison}. B3FA also lands up to $32$ flips, but each flip inverts the sign bit of one \texttt{int8} weight, changing that weight by only $128$ times its channel's scale. As a result, B3FA brings only ResNet8 to random-guess accuracy, and in at most $2\%$ of runs. Under the weaker success criterion of a $10\%$ relative loss of accuracy, DeepHammer outperforms CoarseHammer in four of the $24$ model--DRAM vulnerability pairs, and B3FA in one.

\subsection{End-to-End Attack on a Real System}
\label{sec:end_to_end}

In this section, we run CoarseHammer end-to-end on an Intel Kaby Lake server running Linux 7.0 with two $8$\,GB Samsung DDR4 DIMMs. The CNN victims are PyTorch models that run integer-only inference with \texttt{int32} biases, and the LLM victims keep their normalization weights in \texttt{float32}. Unlike in the DRAM simulations, the attacker cannot choose the frame for each target page, because ASLR moves the target pages between stack positions from load to load.
\begin{figure}[t]
    \centering
    \includegraphics[width=\columnwidth]{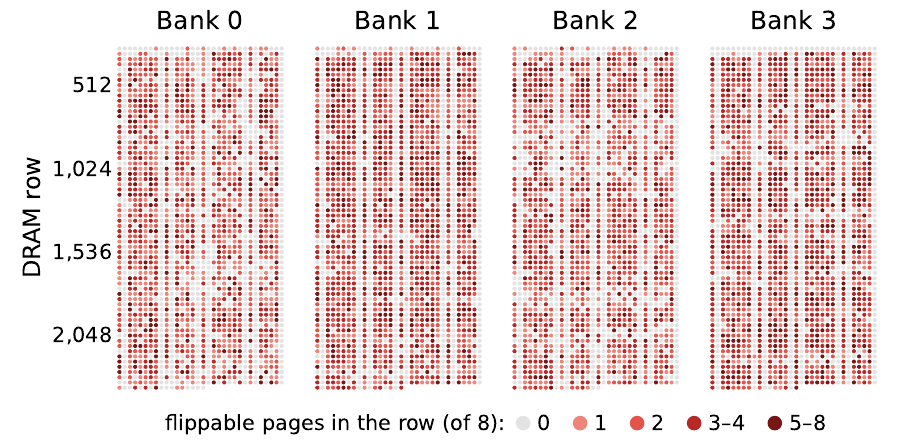}
    \caption{Flippable pages per DRAM row in banks 0–3 after over 256 MiB of physical memory (0x1c23e0000–0x1d29fffff). Each dot is one row, which spans eight 4 KiB pages, and each line holds 32 rows.}
    \label{fig:flip_dist}
\end{figure}

\begin{table*}[t]
\centering
\caption{End-to-end attack on the Kaby Lake server with $256$\,MiB templated. Pages is the mean number of target pages per load, and $S$ is the mean number of target pages per load over the attacker's $N$ frames. Landed flips is the mean number of target bits that flip per load. The last column is the median accuracy over loads with $N=32$.}
\label{tab:end_to_end}
\label{tab:landed}
\label{tab:landed_attack}
\begin{tabular*}{\textwidth}{@{\extracolsep{\fill}}l||rrrrrr||rrrrr||r@{}}
\toprule
\multicolumn{1}{@{}l}{} & & \multicolumn{5}{c}{\textbf{Target pages on the frames, $S$}} & \multicolumn{5}{c}{\textbf{Landed flips per load}} & \textbf{Median accuracy} \\
\cmidrule(lr){3-7}\cmidrule(lr){8-12}\cmidrule(lr){13-13}
\textbf{Victim} & \textbf{Pages} & $N{=}4$ & $N{=}16$ & $N{=}32$ & $N{=}64$ & $N{=}128$ & $N{=}4$ & $N{=}16$ & $N{=}32$ & $N{=}64$ & $N{=}128$ & $N{=}32$ \\
\midrule
ResNet8      & $9.0$   & $1.5$ & $4.6$  & $7.3$     & $8.5$  & $8.6$  & $1.5$  & $2.7$  & $3.0$    & $2.8$  & $3.1$  & $11.1\%$ \\
ResNet18     & $18.5$  & $3.8$ & $10.4$ & $15.6$    & $17.2$ & $17.6$ & $3.6$  & $11.7$ & $13.9$   & $14.2$ & $13.7$ & $10.1\%$  \\
ResNet50     & $50.4$  & $3.8$ & $15.2$ & $27.2$    & $41.7$ & $48.3$ & $5.5$  & $9.6$  & $10.0$   & $10.2$ & $9.8$  & $13.4\%$ \\
DenseNet121  & $124.2$ & $3.2$ & $12.2$ & $23.5$    & $45.6$ & $74.4$ & $0.2$  & $1.0$  & $2.5$    & $3.3$  & $3.9$  & $11.5\%$ \\
MobileNetV1  & $22.4$  & $3.4$ & $9.7$  & $15.4$    & $19.5$ & $21.8$ & $1.3$  & $3.9$  & $5.3$    & $5.3$  & $5.9$  & $50.2\%$ \\
DS-CNN       & $10.5$  & $2.5$ & $7.1$  & $9.5$     & $10.0$ & $9.7$  & $1.9$  & $2.8$  & $2.4$    & $2.6$  & $2.9$  & $9.0\%$  \\
\midrule
Qwen2.5-1.5B & $142$ & $4.0$ & $15.8$ & $31.7$ & $63.0$ & $112.3$ & $19.0$ & $66.9$ & $129.7$ & $247.8$ & $343.8$ & $25.0\%$ \\
Gemma-2-2B   & $289$ & $3.7$ & $15.1$ & $30.3$ & $60.7$ & $121.5$ & $17.9$ & $64.2$ & $124.8$ & $244.9$ & $431.8$ & $25.0\%$ \\
Llama-3.2-3B & $227$ & $3.9$ & $15.6$ & $31.2$ & $62.3$ & $124.6$ & $19.0$ & $66.3$ & $128.5$ & $251.5$ & $444.5$ & $25.0\%$ \\
Phi-4-Mini   & $259$ & $3.9$ & $15.7$ & $31.5$ & $63.0$ & $126.1$ & $19.2$ & $67.0$ & $129.7$ & $254.0$ & $448.8$ & $25.0\%$ \\
\bottomrule
\end{tabular*}
\end{table*}

\paratitle{Address mapping and memory templating}
We reverse-engineered the server's physical-to-DRAM address mapping by sorting physical addresses into banks based on timing row-buffer conflicts~\cite{197211} ($440$-$490$ cycles, versus $280$-$300$ without a conflict). Four XOR functions of the physical address bits select the bank set, and the row index lies in bits $17$ and above. We replaced Blacksmith's address matrices~\cite{jattke_2022_blacksmith} with this mapping, which converts addresses to DRAM coordinates and back, and templated $256$\,MiB of memory across 16 banks to find rows with vulnerable cells and the corresponding aggressor rows and patterns. \autoref{fig:flip_dist} shows how flippable pages are distributed across four banks ($16$\,MiB each). Across these four banks, we find $20{,}088$ flippable bits ($10{,}113$ $0\rightarrow1$ and $9{,}975$ $1\rightarrow0$), a vulnerability rate of $37.4$ flippable bits per million bits.

\paratitle{Reusing a fixed set of frames}
The attacker places its $N$ templated frames at the $N$ stack positions that a target page occupied in the majority of profiling runs, and uses filler frames with no flips in all other positions. The attacker templates DRAM once and reuses the same $N$ frames in every load. We choose these $N$ frames by counting, for each templated frame, the flippable bits at page offsets that hold a target tensor. We give priority to frames whose flips land in the most significant byte of a parameter since these flips change parameters most. When the victim exits, the kernel frees the $N$ frames on the victim's CPU. The attacker keeps each frame's physical neighbors allocated, so the buddy allocator cannot merge the frame into a larger block. The attacker then faults pages on the victim's CPU until it has reclaimed all $N$ frames.

\paratitle{Placement and landed flips}
In all $16.5$K placement runs, every frame occupied its intended stack position and was reclaimed after the load. \autoref{tab:end_to_end} gives, for each $N$, the mean number of target pages on the attacker's frames and of flips that land per load. With $N=4$ frames, at least one flip lands in $98.9\%$ or more of loads for ResNet18, ResNet50, and all four LLMs. Small target tensors are harder to hit because they occupy small portions of a page, so ResNet8, DS-CNN, and DenseNet121 receive a flip in fewer than half of their loads with $N=4$ frames. Placement is cheap and reclaiming each frame on the victim's CPU through the per-CPU free list takes a few microseconds. The end-to-end attack with $N=32$ takes $24$-$32$s, compared to $66$-$96$s reported in prior work~\cite{yao_2020_deephammer}.

\begin{table}[t]
\centering
\caption{Attack success rates for weights quantized below $8$ bits. ResNet
rates are on CIFAR-10 with DoReFa-Net~\cite{2018_shuchang_dorefa}. Llama-3.2-3B uses K-quant~\cite{kawrakow2023kquants}
checkpoints, $1000$ runs each, where success means MMLU accuracy at the random-guess
level ($25\%$) or worse.}
\label{tab:diff_quant}
\def\arraystretch{1.1}
\begin{tabular*}{\columnwidth}{@{\extracolsep{\fill}}l||lcc@{}}
\toprule
\textbf{Model} & \textbf{Quantization} & \textbf{Pristine} & \textbf{Success} \\
\midrule
ResNet8      & W6A6              & $80.0\%$ & $98.1\%$ \\
             & W4A4              & $78.9\%$ & $98.7\%$ \\
\midrule
ResNet18     & W6A6              & $82.4\%$ & $97.1\%$ \\
             & W4A4              & $81.7\%$ & $94.6\%$ \\
\midrule
Llama-3.2-3B & \texttt{Q6\_K}    & $63\%$   & $96\%$   \\
             & \texttt{Q4\_K\_M} & $62\%$   & $96\%$   \\
\bottomrule
\end{tabular*}
\end{table}

\paratitle{Accuracy after the attack}
With $N=32$ frames, the median accuracy over runs is at most $2.2$ percentage points above random-guess accuracy for ResNet50, and all other CNNs are approximately at the random-guess level or worse. In ~\autoref{tab:landed_attack}, we report the accuracies for $N=32$ because it balances the trade-off between attack effectiveness and bit flips landing outside parameter tensors, which may crash the victim process. In this setting, the observed crash rate is less than $3\%$ and only affects the victim process.

\subsection{Different Bit Widths and Fine-Tuning}
\label{sec:persistence}

CoarseHammer selects target tensors by their bit widths and sizes. Changing
parameter values or reducing the precision of other tensors leaves these
selection criteria unchanged. We test whether the selected tensors remain
vulnerable when weights are quantized to different bit widths and when the
victim continues fine-tuning.

\phantomsection
\label{sec:diff_quant}
\paratitle{Different quantization schemes} Changing weight precision leaves the accumulators and biases in \texttt{int32}, so CoarseHammer's target tensors do not change. On Llama-3.2-3B, \texttt{Q6\_K} and \texttt{Q4\_K\_M} formats lower the bit width of the weights but leave the normalization tensors unchanged. CoarseHammer succeeds in $96\%$ of runs in both formats. On ResNet8 and ResNet18 trained with DoReFa-Net~\cite{2018_shuchang_dorefa} at $6$ and $4$ bits, with $78.9\%$-$82.4\%$ pristine accuracy, it brings accuracy to the random-guess level or below in $94.6\%$-$98.7\%$ of runs, as shown in~\autoref{tab:diff_quant}.

\begin{figure}[t]
\centering
\includegraphics[width=\columnwidth]{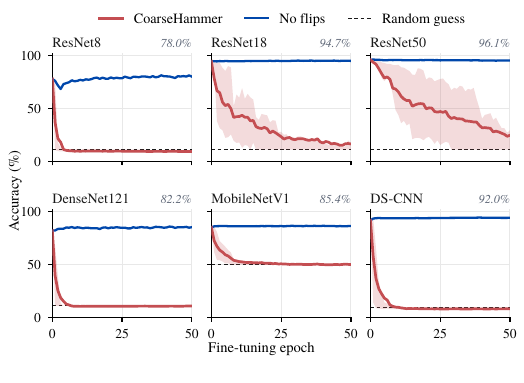}
\caption{Accuracy while the victim fine-tunes for $50$ epochs and CoarseHammer flips $0$-$4$ bits at the start of each epoch. Red lines give the mean over $50$ runs. Blue lines give the mean of five fine-tuning runs without flips, and dashed lines mark random-guess accuracy. The gray percentage beside each model name is its accuracy before fine-tuning.}
\label{fig:finetune}
\end{figure}

\phantomsection
\label{sec:continued_finetuning}
\paratitle{Continued fine-tuning} Fine-tuning during deployment changes the parameter values used for inference, so a bit-flip chain that a bit search computed from the earlier weights becomes stale. CoarseHammer's targets stay the same as fine-tuning does not change the bit width nor the tensor sizes. We fine-tune each of the six CNNs for $50$ epochs. The victim updates the stored tensors with stochastic gradient descent and keeps their quantization scales fixed, so a flipped value persists until an update changes it. At the start of each epoch, CoarseHammer flips $0-4$ bits. Each model gets $50$ attack runs, and five runs that fine-tune without flips. \autoref{fig:finetune} shows the accuracy change with and without flips. Fine-tuning changes up to $40.1\%$ of the \texttt{int8} weights. With CoarseHammer, every run on ResNet8, DenseNet121, MobileNetV1, and DS-CNN reaches random-guess accuracy, after a median of at most $18$ flips. On ResNet18 and ResNet50, the updates recover part of the accuracy between flips, so $92\%$ and $44\%$ of their runs reach random-guess accuracy. Over 50 epochs, mean accuracy falls from $94.7\%$ to $16.4\%$ and from $96.1\%$ to $24.6\%$, respectively.

\subsection{Against BFA Defenses}
\label{sec:CoarseHammer_vs_defs}

We evaluate two categories of BFA defenses. \textit{Range restriction} clips activations that a flipped bias pushes out of range. \textit{Multi-exit models} add classifiers to hidden layers and answer each input from a random exit path, so a fault in one layer changes only the answers of the exits after that layer. In \autoref{sec:defenses}, we discuss other defenses and their implications for CoarseHammer.

\paratitle{Range restriction} Ranger~\cite{9505066} clips the output of each activation layer at the largest value the layer takes on $20\%$ of the training set. We add these clips to the six CNNs, profile their bounds on inputs outside the evaluation examples, and repeat the ideal-placement runs with the same flips. Ranger leaves every model's success rate unchanged. In integer-only inference, each layer already clips its output to its \texttt{int8} range, which the quantization framework calibrates on sample inputs. In 91\% of the activation layers in the models we test, the bounds chosen by Ranger are the minimum and maximum of that range, so an incorrect activation is almost always undetected. A flip in a high-order bias bit drives its channel to such a bound on every input, and clipping keeps it there. Per-channel bounds on every operator output up to the logits, stricter than Ranger's per-layer bounds, leaves CoarseHammer at $96.8\%$-$100\%$ success on five models and at $66.5\%$ on ResNet50, down from $89.8\%$.

\paratitle{Multi-exit models} Aegis~\cite{wang2023aegis} adds an internal classifier (IC) to every residual block and answers each input from a path of randomly picked exits. Aegis trains the ICs on the features of a copy of the model flipped by a gradient-based bit search. Aegis raises the number of flips that BFA~\cite{rakin2019bit} needs to reach random-guess accuracy by up to $38.4\times$. We repeat the ideal-placement runs with the same flips against ResNet18 with Aegis. Every run still reaches random-guess accuracy, after a median of $6$ flips, vs. $4$ without Aegis. Most of CoarseHammer's targets in ResNet18 lie in the first convolution and the first residual stage, whose outputs feed at least eight of the nine exits, so the exits fail together. Even when the ICs train against CoarseHammer's own flips, $99\%$ of runs reach random-guess accuracy, after a median of $10$ flips.

\section{Discussion}
\label{sec:countermeasures}

In this section, we discuss defenses that prevent Rowhammer flips or target-page steering, detect and correct faults, or protect target tensors. We then state CoarseHammer's limitations.

\subsection{Defenses}
\label{sec:defenses}

\paratitle{Preventing flips or the steering}
TRR~\cite{9152631} and counter-based mitigations such as Graphene~\cite{9251863}, TWiCe~\cite{8980327}, and BlockHammer~\cite{9407238} refresh victim rows or throttle the aggressor once its activation count crosses a threshold. Prior work shows that TRR falls against many-sided patterns~\cite{9152631}, and defenses that presuppose adjacent aggressors fall to hammering from two rows away~\cite{281352}. Remapping rows~\cite{9923789, 10070999, 2308.14907} breaks spatial adjacency, and JEDEC has put Per-Row Activation Counting (PRAC) into the DDR5 specification~\cite{jedec_ddr5_2025, 10.1145/3669940.3707278, 10946754, 10.1145/3695053.3731007}. These mitigations require new hardware, and Rowhammer flips have been demonstrated on DRAM chips with ECC~\cite{8835222}. The software detector ANVIL~\cite{2016_Aweke_anvil} needs no new hardware but falls to one-location hammering~\cite{another_flip_2018}. A mitigation that only lowers the number of flippable bits per million helps more against bit searches than against CoarseHammer. Randomizing the PFC's last-in, first-out reuse of freed frames would prevent steering without a hardware change, at a cost in fragmentation and performance. CATT~\cite{2017_brasser_catt} and SoftTRR~\cite{softtrr} protect only kernel memory and page tables. ZebRAM~\cite{2018_Radhesh_zebram} isolates every data row with adjacent guard rows, but can incur up to $50\%$ memory overhead. Citadel~\cite{citadel} physically isolates the memory of different processes, but incurs a mean memory overhead of $7.2\%$. Both ZebRAM and Citadel would stop CoarseHammer as they stop any Rowhammer attack across processes.

\paratitle{Detecting and correcting the fault}

Most defenses against BFAs on parameters protect only the weights~\cite{li2021radar, li2020weightrecon, guan2019inplace, velcicky2026deepncode, nazari2024far}. Weight reconstruction~\cite{li2020weightrecon} assumes that bias and batch-normalization parameters are too small for Rowhammer to target precisely, and in-place ECC~\cite{guan2019inplace} leaves the \texttt{int32} biases unprotected. Inference frameworks, however, load them into DRAM along with other weight tensors, which leaves them exposed to CoarseHammer. HASHTAG~\cite{10.1109/ICCAD51958.2021.9643556} hashes and checks the layers that a bit search attack would target by gradient ranking~\cite{rakin2019bit, yao_2020_deephammer,
coalson2024prisonbreak, rakin2021t}. NeuroPots~\cite{285469} lures gradient-based bit-search attacks into decoy weights protected by checksums. RADAR~\cite{li2021radar} observes that bit-search attacks almost always pick \texttt{int8} MSBs in their chains and protects them. HammerDodger~\cite{gongye2023hammerdodger} and ObfusBFA~\cite{yan2026obfusbfa} periodically move the parameters so that an attacker's precomputed offsets miss, and ObfusBFA assumes that random flips barely affect inference. Since CoarseHammer accepts any flip in its target tensors, it needs neither the bits that these defenses protect nor the offsets that they move. Weight binarization~\cite{9156736} stores the weights in $1$ bit, which raises the mean flips BFA~\cite{rakin2019bit} needs on ResNet-20 from $28$ to $541$. However, binarizing the high-precision parameter tensors that CoarseHammer targets penalizes the accuracy of the model significantly, so it leaves them untouched. BitShield~\cite{chen2025bitshield} covers executables emitted by DNN compilers~\cite{tvm, rotem2018glow}, which leaves out a substantial share of DNNs deployed using general-purpose inference frameworks that load the victim model dynamically, such as PyTorch, TensorFlow, and llama.cpp.

\subsection{Limitations}
\label{sec:limitations}

Bit flips can land in unintended frames due to ASLR. They can corrupt unrelated model parameters, or they can land in the memory allocator's metadata or the inference framework's own objects, which might crash the victim process. In the real-system experiment, using $N=32$ templated frames gives a crash rate below $3\%$ for CNNs. LLMs show a $0\%$ crash rate even with $N=128$ templated frames. This is due to the large size of high-precision parameters in LLMs, which span full pages and thus cover the majority of stray flips.

CoarseHammer requires the target tensors to live in DRAM where an attacker can influence placement and eventually induce flips using Rowhammer. An accelerator that retains high-precision tensors in on-chip SRAM is outside the scope of our Rowhammer implementation. Prior work has shown that on-chip memory can be faulted by a laser beam~\cite{agoyan2010flip, selmke2015precise}, or by a co-tenant on a shared FPGA platform~\cite{rakin2021deepdup}. Further analysis of the vulnerability of these tensors to other fault-injection mechanisms remains open for future work.

\section{Related Work}
\label{sec:related}

\subsection{Bit-flip attacks on parameters}
Invisible Hands~\cite{almalky2025invisible} ranks the weights of a public LLM that shares the victim's architecture on public data and flips the same bits in the victim. CoarseHammer needs neither such proxy weights nor exact bits. T-BFA's~\cite{rakin2021t} overview figure shows bias and batch-normalization parameters as sensitive but on-chip, outside the DRAM it attacks. Hong et al.'s~\cite{hong2019terminal} blind Rowhammer attacker, unaware of where its flips land, still degrades a \texttt{float32} VGG16, since a flip in the top exponent bit of many \texttt{float32} parameters is damaging. However, when the model was quantized to $8$ bits, no such parameters remained. Gongye and Fei~\cite{gongye2024oneflip} inject single-bit flips into quantized models on an FPGA accelerator and find that the scaling factors of quantized layers are single points of failure. Liu et al.~\cite{8203770} changed a single bias to cause a targeted misclassification, and the SBFA~\cite {guo2025sbfa} white-box bit-search attack on \texttt{int8} LLMs also spans floating-point biases and LayerNorm tensors.

\subsection{Bit-flip Attacks on Code}
Other attacks flip bits in code instead of parameters. FrameFlip~\cite{usenix24_onebf_matters} flips bits in the linear-algebra library that the inference framework shares with other processes, and Chen et al.~\cite{comp_models_built_in_expoloits_2025} flip bits in executables produced by DNN compilers~\cite{tvm,rotem2018glow}. Rowhammer also flips unintended bits~\cite{8835222,frodo_2022,pinpoint_2019}, so a code-targeting attack can corrupt instructions in system-wide libraries and affect other programs using them. Bit-search attacks on parameters and on code assume that the attacker can place the vulnerable data on specific chosen DRAM frames and hammer exactly the bit that the search selected~\cite{yao_2020_deephammer,cai_2024_deepvenom,usenix24_onebf_matters,comp_models_built_in_expoloits_2025,iccd24_semiblackbox,wang_2024_groan,almalky2025invisible}.

\subsection{Placing Target Pages on Real Systems}
In prior BFAs that run on real systems, the attacker identifies each target page before steering it onto a templated frame. In DeepHammer~\cite{yao_2020_deephammer}, Don't Knock~\cite{tol_2021_dont_knock}, and Groan~\cite{wang_2024_groan}, the attacker maps the model file and faults its page-cache pages onto templated frames. A file page holds the same bytes wherever ASLR places the file's mapping, so the attacker identifies each target page by its offset in the file. These approaches, however, require the attacker to have storage access to the victim's model file containing the parameters. DeepVenom~\cite{cai_2024_deepvenom} steers the victim's anonymous pages through the PFC just before the victim fills a target weight tensor, whose pages then take the templated frames in order. CoarseHammer's target tensors are instead heap allocations that the victim makes while loading the model, and ASLR changes which heap page holds each of them from load to load.

\section{Conclusion}

Quantization asymmetry leaves small high-precision tensors vulnerable to bit flips. CoarseHammer targets these tensors without parameter values, gradients, or data, and its targets remain vulnerable during fine-tuning. Under ideal placement, CoarseHammer reaches random-guess accuracy in $81.5$-$100\%$ of runs on each of ten evaluated models, with median successful runs requiring at most $24$ flips. CoarseHammer outperforms BFA, DeepHammer, and B3FA on every evaluated CNN at every tested DRAM vulnerability. At $0.01$ flippable bits per million, it succeeds in up to $84\%$ of runs, while all three prior attacks fail.

On a real system, CoarseHammer steers heap-allocated tensors onto vulnerable frames despite ASLR, without accessing the model file, and hammers those frames in every load. With $32$ target frames, the median accuracy over loads is at most $0.3$ percentage points above random-guess accuracy for every model except ResNet50, which drops from $96.1\%$ to $13.4\%$.

\paratitle{Ethics and disclosure} We analyze the vulnerabilities of quantized DNNs to bit flips with the goal of improving the robustness and security of the critical applications that often rely on them. This research involved no human subjects, private data, or sensitive content. All experiments were conducted using open-source models, datasets, and frameworks. Stakeholders include ML users, cloud providers, and hardware vendors.

\bibliographystyle{IEEEtran}
\bibliography{ref}

\end{document}

%% file: figures/overview.tex
\begin{tikzpicture}[x=1pt, y=1pt]
\sffamily
\newcommand{\tw}[1]{{\fontseries{b}\ttfamily #1}}%

\definecolor{ink}{HTML}{111111}
\definecolor{crim}{HTML}{D7263D}
\definecolor{cobalt}{HTML}{0047AB}
\definecolor{gold}{HTML}{F2B200}

\newcommand{\fs}{\fontsize{6.5}{7.5}\selectfont\sansmath}
\newcommand{\fb}{\fontsize{8}{9.5}\selectfont\sansmath}

\tikzset{
  lw1/.style={line width=0.5pt},
  lw2/.style={line width=0.9pt},
  lw3/.style={line width=1.6pt},
  tp/.tip={Latex[length=3.2pt,width=2.8pt]},
  tpb/.tip={Latex[length=4.6pt,width=4.2pt]},
  lab/.style={font=\fs, text=ink, inner sep=0pt},
  blk/.style={draw=ink, lw2, fill=white, minimum width=18pt, minimum height=10pt,
              inner sep=0pt, font=\fs, text=ink},
  sumj/.style={circle, draw=ink, lw2, fill=white, minimum size=7.2pt, inner sep=0pt},
  injj/.style={circle, draw=crim, lw2, fill=white, minimum size=5.6pt, inner sep=0pt},
  clean/.style={draw=cobalt, lw2},
  fault/.style={draw=crim, lw3, line cap=butt},
  dimline/.style={draw=ink, lw1, {Latex[length=3.2pt,width=2.4pt]}-{Latex[length=3.2pt,width=2.4pt]}},
  extl/.style={draw=ink, lw1},
}
\newcommand{\stepb}[4]{%
  \node[circle, fill=ink, text=white, minimum size=9pt, inner sep=0pt,
        font=\fontsize{6.5}{6.5}\selectfont\bfseries] at (#1,#2) {#3};
  \node[font=\fb, text=ink, inner sep=0pt, anchor=west] at ($(#1,#2)+(6.5,0)$) {#4};}
\newcommand{\tags}[2]{%
  \fill[cobalt] ($(#1,#2)+(-5.2,1.6)$) rectangle ++(7,2.4);
  \fill[crim]   ($(#1,#2)+(2.8,1.6)$) rectangle ++(2.4,2.4);}
\newcommand{\sumat}[3]{%
  \node[sumj] (#1) at (#2,#3) {};
  \draw[ink, lw1] ($(#1)+(-2,0)$) -- ($(#1)+(2,0)$) ($(#1)+(0,-2)$) -- ($(#1)+(0,2)$);}

\useasboundingbox (0,0) rectangle (516,152);
\draw[dimline] (0.3,147) -- (515.7,147);
\node[lab, fill=white, inner xsep=2.5pt, font=\fs\bfseries] at (258,147) {OFFLINE};
\draw[dimline] (0.3,72.5) -- (515.7,72.5);
\node[lab, fill=white, inner xsep=2.5pt, font=\fs\bfseries] at (258,72.5) {ONLINE};
\def\ym{106}\def\ys{88}
\stepb{4.5}{137}{1}{Recover architecture}
\stepb{100.5}{137}{2}{Flip spreads downstream}
\node[lab, anchor=east] (lg32) at (350.4,137) {\tw{int32}};
\fill[crim] ($(lg32.west)+(-5.1,-1.2)$) rectangle ++(2.4,2.4);
\node[lab, anchor=east] (lg8) at ($(lg32.west)+(-11.1,0)$) {\tw{int8}};
\fill[cobalt] ($(lg8.west)+(-9.7,-1.2)$) rectangle ++(7,2.4);

\node[blk] (stem) at (14,\ym) {conv};
\node[blk] (s1a)  at (45,\ym) {conv};
\node[blk] (s1b)  at (70,\ym) {conv};
\node[blk] (s2a)  at (137,\ym) {conv};
\node[blk] (s2b)  at (161.5,\ym) {conv};
\node[blk] (s3a)  at (217,\ym) {conv};
\node[blk] (s3b)  at (241.5,\ym) {conv};
\node[blk, minimum width=16pt] (pool) at (290,\ym) {pool};
\node[blk, minimum width=14pt] (fc)   at (312,\ym) {FC};
\node[blk] (sc2) at (157,\ys) {1$\times$1};
\node[blk] (sc3) at (237,\ys) {1$\times$1};
\foreach \x in {14,45,70,137,161.5,217,241.5,312} {\tags{\x}{111}}
\foreach \x in {157,237} {\tags{\x}{93}}
\node[lab, text=crim, anchor=south] at (74,116.4) {$T$};

\node[injj] (J) at (87,\ym) {};
\draw[crim, lw1] ($(J)+(-1.6,0)$) -- ($(J)+(1.6,0)$) ($(J)+(0,-1.6)$) -- ($(J)+(0,1.6)$);
\sumat{a1}{112}{\ym}
\sumat{a2}{192}{\ym}
\sumat{a3}{272}{\ym}

\draw[clean, -tp] (0,\ym) -- (stem.west);
\draw[clean] (stem.east) -- (29,\ym);
\draw[clean, -tp] (29,\ym) -- (s1a.west);
\draw[clean, -tp] (s1a.east) -- (s1b.west);
\draw[clean, -tp] (s1b.east) -- (J.west);
\draw[clean, -tp] (29,\ym) -- (29,\ys) -- (112,\ys) -- (a1.south);
\fill[cobalt] (29,\ym) circle (1.5);

\draw[fault, -tpb] (J.east) -- (a1.west);
\draw[fault] (a1.east) -- (122,\ym);
\draw[fault, -tpb] (122,\ym) -- (s2a.west);
\draw[fault, -tpb] (s2a.east) -- (s2b.west);
\draw[fault, -tpb] (s2b.east) -- (a2.west);
\draw[fault, -tpb] (122,\ym) -- (122,\ys) -- (sc2.west);
\draw[fault, -tpb] (sc2.east) -- (192,\ys) -- (a2.south);
\draw[fault] (a2.east) -- (202,\ym);
\draw[fault, -tpb] (202,\ym) -- (s3a.west);
\draw[fault, -tpb] (s3a.east) -- (s3b.west);
\draw[fault, -tpb] (s3b.east) -- (a3.west);
\draw[fault, -tpb] (202,\ym) -- (202,\ys) -- (sc3.west);
\draw[fault, -tpb] (sc3.east) -- (272,\ys) -- (a3.south);
\draw[fault, -tpb] (a3.east) -- (pool.west);
\draw[fault, -tpb] (pool.east) -- (fc.west);
\draw[fault, -tpb] (fc.east) -- (330,\ym);
\fill[crim] (122,\ym) circle (2.1) (202,\ym) circle (2.1);
\node[lab, anchor=west] at (332.5,\ym) {logits};

\draw[crim, lw2] (87,124.2) circle (4.3);
\draw[crim, lw2, -{Latex[length=2.4pt,width=2.4pt]}] (87,127.3) -- (87,121.2);
\draw[crim, lw2, -tp] (87,119.9) -- (J.north);
\node[font=\fb, text=crim, inner sep=0pt, anchor=west] at (93.6,124.2) {bit flip};
\stepb{368.5}{137}{3}{Template DRAM}
\def\tx{369}\def\tq{6.8}\def\ty{116.5}
\foreach \r in {0,...,3} {
  \draw[ink, lw1] (\tx-3,\ty-\r*7) -- (\tx+19*\tq+3,\ty-\r*7);
  \foreach \c in {0,...,19} {
    \draw[ink, lw1, fill=white] (\tx+\c*\tq-1.5,\ty-\r*7-1.5) rectangle ++(3,3);}}
\foreach \r/\c in {0/3,1/15,2/8,3/12} {
  \fill[crim] (\tx+\c*\tq-1.75,\ty-\r*7-1.75) rectangle ++(3.5,3.5);}
\draw[crim, lw2] (406.4,99.1) rectangle (440.4,105.9);
\stepb{4.5}{60}{4}{Steer target page}
\filldraw[fill=white, draw=ink, line width=0.5pt] (0.25,39) rectangle (32.25,49);
\node[lab, font=\fs, anchor=base] at (16.25,41.67) {planter};
\filldraw[fill=white, draw=ink, line width=0.5pt] (167.75,39) rectangle (199.75,49);
\node[lab, font=\fs, anchor=base] at (183.75,41.67) {victim};
\draw[draw=ink, line width=0.5pt] (82,35) -- (82,5) -- (118,5) -- (118,35);
\filldraw[fill=white, draw=crim, line width=0.9pt] (84,7) rectangle (116,12);
\fill[fill=crim] (98.6,8.1) rectangle (101.4,10.9);
\filldraw[fill=white, draw=ink, line width=0.5pt] (84,14) rectangle (116,19);
\filldraw[fill=white, draw=ink, line width=0.5pt] (84,21) rectangle (116,26);
\filldraw[fill=white, draw=ink, line width=0.5pt] (84,28) rectangle (116,33);
\node[lab, font=\fs, anchor=base] at (100,41.67) {PFC list};
\draw[draw=ink, line width=0.5pt, -tp] (32.5,44) -- (85.5,44) -- (85.5,33.25);
\draw[draw=ink, line width=0.5pt, -tp] (114.5,33.25) -- (114.5,44) -- (167.5,44);
\filldraw[fill=crim, draw=crim, line width=0.5pt] (167.75,7) rectangle (199.75,12);
\filldraw[fill=white, draw=ink, line width=0.5pt] (167.75,14) rectangle (199.75,19);
\filldraw[fill=white, draw=ink, line width=0.5pt] (167.75,21) rectangle (199.75,26);
\filldraw[fill=white, draw=ink, line width=0.5pt] (167.75,28) rectangle (199.75,33);
\draw[draw=ink, line width=0.5pt] (79.6,14) -- (78,14) -- (78,33) -- (79.6,33);
\node[lab, font=\fs, anchor=base east] at (76,21.17) {$n{-}1$ fillers};
\node[lab, font=\fs, anchor=base east, text=crim] at (76,7.17) {$F_{i}$};
\node[lab, font=\fs, anchor=base east, text=crim] at (163.75,7.17) {target page};
\draw[draw=crim, line width=0.9pt, -tp] (199.75,9.5) -- (221.22,9.5) -- (221.22,22.5) -- (242.7,22.5);
\stepb{244.5}{60}{5}{Hammer}
\begin{scope}[shift={(-145,-70.5)}]
\def\cx{392}\def\cp{8.6}
\foreach \c in {0,...,5} {\draw[ink, lw1] (\cx+\c*\cp,75.5) -- (\cx+\c*\cp,122.5);}
\draw[ink, lw1] (385,119) -- (441,119);
\draw[ink, lw2] (385,106) -- (441,106);
\draw[ink, lw2] (\cx-4.3,93) -- (441,93);
\draw[ink, lw2] (385,80) -- (441,80);
\foreach \yy in {106,80} {
  \draw[gold, lw3, line join=miter] (441,\yy) -- ++(3,0) -- ++(0,4.5) -- ++(3,0) -- ++(0,-4.5)
    -- ++(3,0) -- ++(0,4.5) -- ++(3,0) -- ++(0,-4.5) -- ++(3,0) -- ++(0,4.5) -- ++(3,0)
    -- ++(0,-4.5) -- ++(2.5,0);}
\foreach \yy in {119,106,93,80} {\foreach \c in {0,...,5} {
  \draw[ink, lw1, fill=white] (\cx+\c*\cp-1.6,\yy-1.6) rectangle ++(3.2,3.2);}}
\fill[crim] (\cx+3*\cp-1.8,93-1.8) rectangle ++(3.6,3.6);
\draw[crim, lw2] (\cx-4.3,89.4) rectangle (\cx+4*\cp+4.3,96.6);
\node[lab, text=crim, anchor=west] at (443.5,93) {0$\to$1};
\node[lab, anchor=west] at (467,106) {aggressor};
\node[lab, anchor=west] at (467,93) {victim row};
\node[lab, anchor=west] at (467,80) {aggressor};
\end{scope}
\stepb{372.5}{60}{6}{Accuracy falls to random guess}
\def\sx{392}\def\sw{23.394}\def\sy{12}\def\sh{0.5}
\draw[ink, lw1] (\sx,\sy) rectangle (510,\sy+80*\sh);
\foreach \k/\l in {0/0,1/1,1.585/2,2.322/4,3.170/8,4.087/16,5.044/32} {
  \draw[ink, lw1] (\sx+\k*\sw,\sy) -- ++(0,2);
  \node[lab, anchor=north] at (\sx+\k*\sw,\sy-1.5) {\l};}
\node[lab, anchor=north east] at (\sx-5,\sy-1.5) {flips};
\draw[ink, lw1, dashed] (\sx,\sy+10*\sh) -- (510,\sy+10*\sh);
\draw[ink, lw2, densely dashed] (\sx,\sy+78*\sh) -- (\sx+3.170*\sw,\sy+74.9*\sh)
  -- (\sx+4.087*\sw,\sy+71.1*\sh) -- (\sx+5.044*\sw,\sy+65.5*\sh);
\draw[crim, lw3, line join=round] (\sx,\sy+78*\sh)
  \foreach \k/\a in {1/40.1,1.585/23.8,2/15.8,2.322/13.1,2.585/11.3,2.807/10.5,3/10.0,3.170/9.9,4.087/9.6,5.044/9.6}
  { -- (\sx+\k*\sw,\sy+\a*\sh) };
\foreach \k/\a in {0/78,1/40.1,1.585/23.8} {\fill[crim] (\sx+\k*\sw,\sy+\a*\sh) circle (1.6);}
\node[lab, anchor=east] at (\sx-4.5,\sy+78*\sh) {78.0\,\%};
\node[lab, text=crim, anchor=south west] at (\sx+1*\sw+2,\sy+40.1*\sh+1) {40.1};
\node[lab, text=crim, anchor=south west] at (\sx+1.585*\sw+2,\sy+23.8*\sh+1) {23.8};
\node[lab, text=crim, anchor=west] at (\sx+2.75*\sw,\sy+38*\sh) {CoarseHammer};
\node[lab, anchor=south east] at (508,\sy+10*\sh+1.5) {random guess 10\,\%};
\node[lab, anchor=north east] at (508,\sy+65.5*\sh-2) {random baseline 65.5};
\end{tikzpicture}